\documentclass[12pt,a4paper]{article}
\usepackage[utf8]{inputenc}
\usepackage[english]{babel}
\usepackage{amsmath}
\usepackage{amsfonts}
\usepackage{amssymb}
\usepackage{graphicx}
\usepackage[affil-it]{authblk}

\title{Does the Wolfenstein form work for the leptonic mixing matrix?}
\author{G Rajasekaran \thanks{Email: graj@imsc.res.in}}

\affil{The Institute of Mathematical Sciences, Chennai \\ and \\ Chennai Mathematical Institute, Chennai}
\date{}

\begin{document}

\maketitle
\begin{abstract}
Starting with the Wolfenstein form for the leptonic mixing matrix we show that renormaliztion group evolution brings that to the observed large mixing at low energies.
\end{abstract}

\section{Introduction and Summary}

It is wellknown that the Cabibbo-Kobayashi-Markawa (CKM) is approximately of the Wolfenstein form \cite{Wolfenstein:1983yz}
$$
\left(
\begin{array}{ccc}
1 & O(\lambda) & O(\lambda^3) \\
O(\lambda) & 1 & O(\lambda^2) \\
O(\lambda^3) & O(\lambda^2) & 1 \\
\end{array}
\right)
$$
where $\lambda$ is a small parameter of the order of Cabibbo angle.  This is highly suggestive of perturbative inter-generational mixing.  To the zeroth order of perturbation, the mixing matrix is a unit matrix.  Generation 1 and 2 mix in first order of $\lambda$, 2 and 3 mix in second order while 3 and 1 mix in third order.  Such a structure is a very important hint towards a theory of genertions.  If that is correct, the Wolfenstein form should be valid for the leptonic mixing also.  But that is far from true.  Leptonic mixing angles $\theta_{12}$ and $\theta_{23}$ are large while $\theta_{31}$ is small.

Here we point out that this mistery can be solved, once it is recognized that the theory of generations that leads to the Wolfenstein perturbative structure may be a high-sclae theory and so the Wolfenstein structre for both quarks and leptons is expected to be valid only at the high energy scale.  Renormalization group must be used to evolve the mixing matrix down to the low energy scale.  While the qwark mixing matrix does not change much, the leptonic mixing matrix changes drastically because of the quasi-degeneracy of neutrino masses, resulting in the observed large $\theta_{12}$ and $\theta_{23}$, and small $\theta_{31}$.  We explicitly demonstrate this result.

\section{Neutrino Mixing Matrix and Masses}
The Pontecorvo-Minakata-Nakano-Sakata(PMNS) neutrino mixing matrix and the neutrino masses as determined by the various osillation experiments are given here.  The mixing matrix

$$U = \left( \begin{array}{c c c}
c_{12}c_{13} & s_{12}c_{13} & s_{13} \\
-s_{12}c_{23}-c_{12}s_{23}s_{13} & c_{12}c_{23}-s_{12}s_{23}s_{13} & s_{23}c_{13}\\
s_{12}s_{23}-c_{12}c_{23}s_{13} & -c_{12}s_{23}-s_{12}c_{23}s_{13} & c_{23}c_{13}
\end{array} \right)$$

where $s_{12} = \sin \theta_{12} $, $c_{12} = \cos \theta_{12}$, $etc.$ and
\begin{align*}
\theta_{12} &\approx 30^0 \\
\theta_{23} &\approx 45^0 \\
\theta_{31} &\approx 9^0
\end{align*}
The neutrino mass differences are,
\begin{align*}
\delta m^2_{21}  &\equiv m^2_{2} - m^2_{1} \simeq 7 \times 10^{-5} eV^2 \\ 
\delta m^2_{32}  &\equiv m^2_{3} - m^2_{2} ; ~ |\Delta m^2_{32}| \simeq 2 \times 10^{-5} eV^2 \\ 
\end{align*}
we ignore the undetermined CP violating Dirac phase and the two Majorana phases in this work.

\section{How RG evolution solves the large angle problem}
At high scale,both the CKH matrix and PMNS matrix are assumed to be of the Wolfenstein form 
\begin{equation}
U = \left( \begin{array}{ccc}
1&0&0\\
0&1&0\\
0&0&1\\
\end{array} \right) + \left(
\begin{array}{ccc}
0 & O(\lambda) & O(\lambda^3) \\
O(\lambda) & 0 & O(\lambda^2) \\
O(\lambda^3) & O(\lambda^2) & 0 \\
\end{array}
\right)
\label{eq:1}
\end{equation}
More correctly, write $U$ in terms of $\theta_{12}$, $\theta_{23}$, and $\theta_{31}$, with $\sin\theta_{12}\approx \lambda$, $\sin\theta_{23}\approx \lambda^2$, $\sin\theta_{12}\approx \lambda^3$.  Use renormalzation group equation to evolve $U$ to low scales.

We find that the CKM matrix does not change much, but the PMNS matrix change dramatically because of the quasi-regenerate nature of the neutrino masses.

Earlier in a series of papers (\cite{Mohapatra:2003tw} - \cite{Abbas:2015vba}), we had derived the consequences of assuming that the CKM and PMNS matrix are unified at high scale:
$$U_{CKM} = U_{PMNS}$$
Using this assumption of high scale mixing unification and RG evolution we could explain the observed neutrino mixing angles and mass-differences observed at low energies.

In the present work we are taking the point of view that unification is \underline{not} necessary.  What is needed is only the Wolfenstien structure (Eq.\ref{eq:1}) with $\lambda$ small (0.1 to 0.3).  RG evolution then magnifies the angles.

\section{Renormalization group evolution}

We shall be brief.  More details can be obtained from the earlier paper.

We assume Majorana neutrinos which get their tiny masses through the standard type I seasaw mechanism.  The renormalization group (RG) evolution equation for the neutrino mass matrix $M_{\nu}$ is

\begin{equation}
\label{eq:2}
16 \pi^2 \dfrac{dM_\nu}{dt} = \left\lbrace -\left( \frac{6}{5} g_1^2 + 6 g_2^2  \right) + Tr\left( 6 Y_U Y_U^\dagger \right) \right\rbrace M_\nu \\\  
+ \frac{1}{2} \left\lbrace \left( Y_E Y_E^\dagger \right) M_\nu + M_\nu \left( Y_E Y_E^\dagger \right) \right\rbrace
\end{equation}
where $t=\ln \mu$, $\mu$ being the scale parameter, $g_{1}$ and $g_{2}$ are the U(1) and SU(2) gauge coupling constants.  $Y_U$ and $Y_E$ are the Yukawa coupling matrices for the up-quark and charged leptons.  We take 

\begin{equation}
\label{eq:3}
Y_U Y_U^\dagger \approx \left( \begin{array}{ccc} 0&& \\ &0& \\ & & h^2_{t}
\end{array} \right), ~ Y_E Y_E^\dagger \approx \left( \begin{array}{ccc} 0&& \\ &0& \\ & & h^2_{\tau}
\end{array} \right)
\end{equation}

where $h_t$ and $h_\tau$ are the higgs coupling of the top and tau respectively.  We have to divide $h^2_t$ by $\sin^2 \beta$ and $h^2_\tau$ by $\cos^2 \beta$ for MSSM, where $\tan \beta = \frac{<\phi^0_u>}{<\phi^0_d>}$.

The PMNS matrix $U$ diagonalizes the mass matrix $M_{\nu}$ in the flavour basis: $U^T M_\nu U = diag (m_1,m_2, m_3)$.  The RG equation for the mass eigenvalues and the mixing angles are
\begin{align}
\label{eq:4to7}
\dfrac{dm_i}{dt} &= - 2F_\tau m_i U^2_{\tau i} - m_i F_U ~(i = 1, 2, 3) \\
\dfrac{ds_{23}}{dt} &= - F_\tau c^2_{23} (-s_{12} U_{\tau 1} D_{31} + c_{12} U_{\tau 2} D_{32} ) \\
\dfrac{ds_{13}}{dt} &= - F_\tau c_{23} c^2_{13} (c_{12} U_{\tau 1} D_{31} + s_{12} U_{\tau 2} D_{32} ) \\
\dfrac{d2_{12}}{dt} &= -F_{\tau} c_{12} (c_{23} s_{13} s_{12} U_{\tau 1} D_{31} - c_{23}s_{13}c_{12}U_{\tau 2} D_{32} + U_{\tau 1} U_{\tau 2} D_{21} )
\label{eq:7}
\end{align}

where $D_{ij} = \frac{m_i+m_j}{m_i-m_j}$; $i \neq j$ and $F_{\tau}$ and $F_\nu$ are in Table \ref{tab:tab1}.
\begin{table}
\centering
\caption{$F_{\tau}$ and $F_{\nu}$ in MSSM and SM}
\label{tab:tab1}
\begin{tabular}{|c|c|c|}
\hline
& $F_\tau$ & $F_U$ \\
\hline
MSSM & $-\frac{h^2_{\tau}}{16 \pi^2 \cos^2_{\beta}}$ & $\frac{1}{16\pi^2} \left( \frac{6}{5} g_1^2 + 6 g_2^2 - \frac{6h_t^2}{\sin^2 \beta}\right) $ \\
\hline
SM & $\frac{3h_\tau^2}{32\pi^2}$ & $\frac{1}{16\pi^2} \left( 3 g_2^2 - 2\lambda - 6h^2_t - 2h^2_\tau \right)$ \\
\hline
\end{tabular}
\end{table}

In MSSM, $F_\tau$ is enhanced by a factor $ \sim 10^{3}$ for $\tan\beta\approx {50}$ as compared to its value in SM.  So the rapid evolution of angles that is needed can be attribured to SUSY.  For quasidegenerate neutrino masses, $D_{ij} \rightarrow \infty$.  This also contributes to rapid evolution of the angles, since the right hand side of the equation for $\frac{ds_{ij}}{dt}$ contain $D_{ij}$.

At high scale the following approximation are valid:
$$s_{12} \sim \lambda \sim 0.2 ;~ s_{23} \sim O(\lambda^2) \sim 0.04 ;~ s_{31} \sim O(\lambda^3) \sim 0.008  ;~ U_{\tau 1} \sim O(\lambda^3) ;~ U_{\tau 2} \sim O(\lambda^2)$$
This allows us to write the following approximate evolution equation which helps us to understand our results in a simple way:
\begin{align}
\label{eq:8to9}
\frac{ds_{23}}{dt} &\sim \lambda^2 F_\tau D_{32} \\
\frac{ds_{13}}{dt} &\sim \lambda^3 F_\tau (D_{32}+D_{31}) \\
\frac{ds_{12}}{dt} &\sim \lambda^5 F_\tau D_{21}
\end{align}
We must also remember $|D_{31}| \approx |D_{32}| << |D_{21}|$.  Hence $s_{23}$ evolves fast, faster than $s_{12}$.For $\dfrac{ds_{12}}{dt}$, the smallness of $\lambda^{5}$ is compensated by the largeness of $D_{21}$.  $\dfrac{ds_{12}}{dt}$ remains small because of $\lambda^{3}$.

\begin{figure}
	\includegraphics[width=\linewidth]{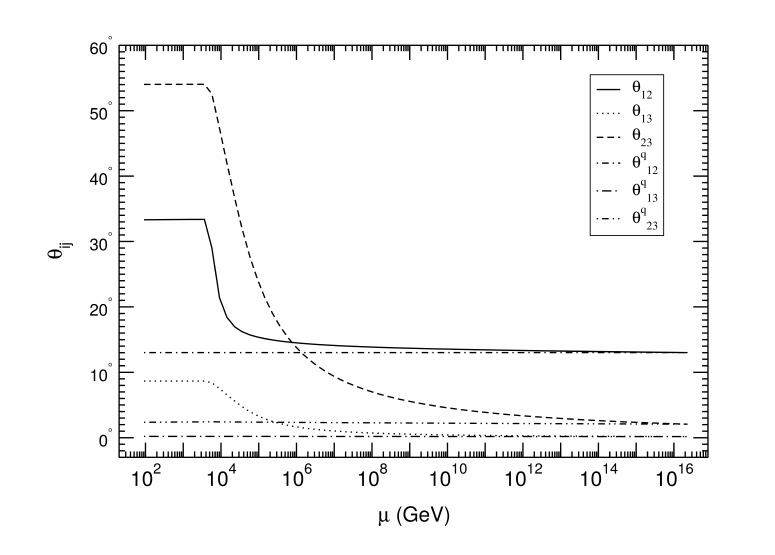}
	\caption{RG evolution of mixing angles}
	\label{fig:rgmix}
\end{figure}

\begin{figure}
	\includegraphics[width=\linewidth]{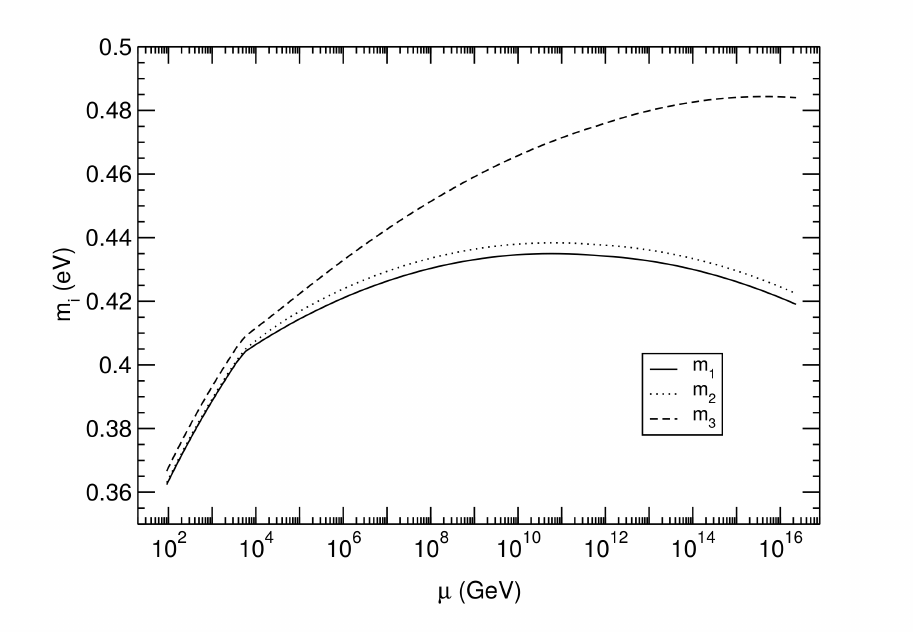}
	\caption{RG evolution of neutrino masses}
	\label{fig:rgmass}
\end{figure}

The results of the full RG evolution eqs (\ref{eq:4to7}) to (\ref{eq:7}) are shown in Figures \ref{fig:rgmix} and \ref{fig:rgmass}.  We see that while the neutrino masses do not evolve much, the neutrino angles evolve very rapidly.  Fig \ref{fig:rgmix} also shows the evolution of the quark angles $\theta^q$ which do not evolve much.  We have taken the SUSY scale as about 10 Tev (as an example).  Starting with all the angles satisfying the Wolfenstein ansatz at high scale, $\theta_{12}$, $\theta_{23}$ and $\theta_{31}$ reach about $30^0, 45^0$ and $10^0$  respectively at low scales.  It is important to note that for the large magnification of the angles, SUSY is essential.

To sum up, we have shown that just like in the quark sector, in the leptonic sector also the mixing angles do satisfy Wolfenstien ansatz at high scale and because of the RG evolution they increase at low engeries, do watch the experimental values.

A final remark: Since both quarks and leptons obey the Wolfenstein ansatz, the real reason for the ansatz must be found.
\section{Acknowledgement} I thank Rahul Srivastava for discussions.

\end{document}